\documentclass[aps,pre,reprint]{revtex4-1}
\usepackage [utf8]{inputenc}
\usepackage[T1]{fontenc}
\usepackage{lmodern}
\usepackage{amsmath}
\usepackage{siunitx}
\DeclareSIUnit{\calorie}{cal}
\usepackage{hyperref}
\usepackage{mathrsfs}
\usepackage{multirow}
\usepackage{texdraw}
\usepackage{color}
\usepackage{epstopdf}
\usepackage[font=small,labelfont=bf]{caption}

\usepackage{graphicx}

\usepackage{color}

\begin{document}
\thispagestyle{empty}

\title{Can we describe charged nanoparticles with electrolyte theories? Insight from mesoscopic simulation techniques}      

\author{Vincent Dahirel}
\email{vincent.dahirel@sorbonne-universite.fr}
\affiliation{Sorbonne Universit\'e, CNRS, Physico-chimie des \'electrolytes et nano-syst\`emes interfaciaux, PHENIX, F-75005 Paris, France} 

\author{Olivier Bernard}
\email{olivier.bernard@sorbonne-universite.fr}
\affiliation{Sorbonne Universit\'e, CNRS, Physico-chimie des \'electrolytes et nano-syst\`emes interfaciaux, PHENIX, F-75005 Paris, France} 

\author{Marie Jardat}
\email{marie.jardat@sorbonne-universite.fr}

\affiliation{Sorbonne Universit\'e, CNRS, Physico-chimie des \'electrolytes et nano-syst\`emes interfaciaux, PHENIX, F-75005 Paris, France}

\begin{abstract}
Electrolyte theories enable to describe the structural and dynamical properties of simple electrolytes in solution, such as sodium chloride in water. Using these theories for aqueous solutions of charged nanoparticles is a straightforward route to extract their charge and size from experimental data. Nevertheless, for such strongly asymmetric electrolytes, the validity of the underlying approximations  have never been properly challenged with exact simulation results. In the present work, well established mesoscopic numerical simulations are used to challenge the ability of advanced electrolyte theories to predict the electrical conductivity of suspensions of charged nanoparticles, in the salt-free case. The theories under investigation are based on the Debye-Fuoss-Onsager treatment of electrolyte transport. When the nanoparticles are small enough (about one nanometer large), the theoretical results agree remarkably well with the simulation ones, even in the high concentration regime (packing fraction of in nanoparticles larger than 3$\%$ ). Strikingly, for highly charged nanoparticles, the theory is able to capture the non-monotonic variation of the ratio of the electrical conductivity to its value at infinite dilution (ideal value) as a function of the concentration. 
However, the tested theories fail to describe the conductivity of suspensions containing larger nanoparticles ({\em e.g.} of diameter $4$~nm). Finally, only small charged nanoparticles can be considered as {\em ions}, as far as electrolyte theories are concerned.  
\end{abstract}



\maketitle

\section{Introduction}

All aqueous solutions contain electrolytes. Electrolyte theories are almost one hundred year old, starting from the Debye-H\"uckel theory of electrostatic screening in 1923 \cite{Debye23}. The range of validity of the Debye-H\"uckel theory is restricted to very dilute solutions, and therefore it is not relevant for most systems of interest. Electrolyte theories have been developing continuously, in parallel to the design of experimental setups to study electrolyte solutions, and motivated by the need to understand more and more diverse types of electrolytes, {\em e. g.} in batteries, in soils, in living cells. The most popular model is the primitive model of electrolyte solutions. It works well to describe equilibrium properties like the evolution of the osmotic coefficients with electrolyte concentration\cite{Dufreche02,Molina11}, or transport properties such as the electrical conductivity\cite{JardatJCP99,dufrecheJPCB05}. It describes the fluid as a gas of charged hard spheres immersed in a continuum characterised by its dielectric constant. This model can be derived using the McMillan Mayer theory\cite{mcmillan}. The crudest way to account for the finite size of individual ions is the extended Debye-H\"uckel theory, which is analogous to the DLVO theory of colloids, with a charge renormalisation $Z^*=Z e^{\kappa a}/(1+ \kappa a)$ due to the absence of screening charges at a distance lower than the radius $a$ from the center of each ion\cite{hansen}. A more elaborate and consistent way of deriving properties of the primitive model have been built since 1970 through the use of integral equations from statistical mechanics\cite{hansen,blum}. The analytic solution for the structure of an electrolyte using the Ornstein-Zernike equation with the Mean Spherical Approximation (MSA) closure has been derived by Waisman, Lebowitz and Blum \cite{Waisman72,blum,blum2}. 

When it comes to experiments, the dynamics of electrolytes is much easier to measure than their structural properties. The electrical conductivity is used for routine measurements of electrolyte concentration, such as in classical practicals for physical chemistry students. Conductivity cells can be built to measure conductivity with great precision. If the temperature of the experimental cell  is correctly controlled,
one is usually able to measure the electrical conductivity with a relative precision lower than 0.1~\%\cite{livrekunz}. When it comes to asymmetric electrolytes including charged nanoparticles, other dynamical quantities, such as the diffusion coefficient \cite{DahirelSoftMatt}, the Ionic Vibration Potential\cite{Pusset2015}, or the electrophoretic mobility\cite{Lucas2014} can be determined. The analysis of such dynamic measurement can  lead to the charge and the size of the particles if an accurate theory is used to analyze the  data\cite{serge,DahirelSoftMatt}. This offers an affordable alternative to very expensive structural studies of the same system. It can also lead to define the speciation of associated colloids\cite{turqcondasso,DurandVidalJPC06} or to predict the stability of a charge-stabilized system\cite{Lucas2014}. 

Even at very low concentrations, the electrical conductivity of electrolyte solutions strongly deviates from its ideal value, that is the value we would obtain without any interaction between ions, {\em i.e. } at infinite dilution.    The scaling of 
the electrical conductivity
as the square root of the concentration has been first derived from a theoretical point of view by Debye, Fuoss, Onsager and Falkenhagen \cite{onsa32,onsa45,onsa57}. Their prediction agrees with Kohlrausch empirical limiting laws for the variation of the conductivity, but it is limited to a concentration range below $10^{-2}$ mol.L$^{-1}$ in water. The theory includes two main deviations from ideality: hydrodynamic interactions between ions, and the electrostatic force arising from the motion of ions. Both these forces are proportional to the applied electric field, within a linear response approach, and are usually called corrective forces. These theories have been improved by using more sophisticated pair correlation functions than those of Debye-H\"uckel's\cite{livrekunz}.

In 1992, Bernard and coworkers extended the scope of application of the Fuoss-Onsager (FO) theory by using equilibrium distribution functions deduced from the  Ornstein-Zernike (OZ) integral equations. Solutions of these equations, obtained either with the Mean Spherical Approximation (MSA) or with the Hypernetted chain (HNC) closure relation \cite{HNC,hansen}, were used to compute the electrical conductivity of electrolyte solutions containing two ions of opposite charges\cite{bernard2}. Typical 1-1 electrolytes (NaCl, KBr, KCl) are well described by this theory below 1 mol.L$^{-1 }$, without any fitting parameters (the size are the crystallographic radii, and the input diffusion coefficient at infinite dilution  $D^{\circ}$ is obtained by extrapolating conductivity data at infinite dilution). The use of the HNC closure equation instead of MSA enables to describe 2-1 electrolytes as well (BaCl$_2$)\cite{bernard2}. In 1995, the theory was extended to associated systems, introducing neutral pairs of ions through an equilibrium constant, computed using electrostatic arguments\cite{turqcondasso}. The agreement with experiments was very good for MgSO$_{4}$, and for LiBr in acetonitrile, with a unique fitting parameter: the minimum interionic distance that is used to estimate the association constant. This success encouraged the use of conductivity as a first choice tool to analyze quantitatively ionic association. The same year, another work extended the conductivity to a mixture of three ionic species\cite{serge}. In 2007, the theory was extended to mixture of any number of ions \cite{VanDamme07}. Later, expressions derived from this work were applied to describe simple electrolyte mixtures containing four species\cite{Roger2009} and buffer solutions \cite{Bernard18}. In 2015, a new way of computing the corrective forces iteratively was suggested, and adapted to the computation of the ionic vibration potential (IVP)\cite{Pusset2015,Gourdin15}. This is the latter implementation that we present in the current paper, in which we specify formulas for the electrolyte conductivity of electrolyte solutions containing two ions. In what follows, we denote by Self-consistent Fuoss-Onsager theory (SFO) this version of the transport theory. 

Transport theory of electrolyte solutions based on the Fuoss-Onsager equation have been extensively validated for simple electrolytes for which good agreement has been obtained with Brownian dynamic simulations \cite{Dufreche02,Jardat00}. Apart from the simulations, other recent theoretical approaches have been developed such as the mode coupling theory\cite{Chandra99,Chandra00} or generalized Langevin theory\cite{Yamaguchi07,Yamaguchi09} which allows to consider  transport measurements dependent on the frequency. At zero frequency and at low concentrations these
approaches also lead to the Debye-H\"uckel-Onsager limiting law \cite{Chandra99}.  Moreover, transport theory based on the Fuoss-Onsager equation have also been used for small charged colloidal particles ({\em e. g.} in Lucas {\em et al} \cite{Lucas2014}), even if in this particular range of ion size, the theory was not validated. Actually, to describe the case of charged nanoparticles,  it is difficult to choose the right conceptual framework: Are suspensions of charged nanoparticles still electrolyte solutions, or should they be viewed through the lens of colloidal theories ? Theories of colloidal conductivity are adapted to the description of suspensions in which the length scales relative to the colloid and the ions of the supporting electrolyte are very different (see for instance \cite{CuquejoJPC06,Chassagne13} for static,  and \cite{Chassagne01,Chassagne03} for frequency dependent measurements). For dispersions containing colloidal particles with a nanometric size -a radius smaller than 10~nm-, such theories do not seem to be relevant. Such cases are very interesting as many systems of interest include asymmetric electrolytes, with charged nanoparticles whose size lies between 1 and 10 nanometers, like proteins, micelles, or inorganic particle. Moreover, conversely to colloidal suspensions, the contribution of the macroion to the electrical conductivity is important, as the ideal conductivity scales with  $\sum_i C_iZ_i^2/R_i$, where $C_i$, $Z_i$ and $R_i$ are, respectively, the concentration, the charge and the radius of the species $i$. In these cases, the measurement of the electrical conductivity can be used to determine the charge of the particles using a fitting procedure, conversely to the case of typical colloidal particles of larger size \cite{Perger07,Medos15}. 

For example, the Fuoss-Onsager theory combined with the MSA\cite{bernard2,serge} has been used with the hard sphere radii and charges (and sometimes the diffusion coefficient) as fitting parameters. For some systems, a unique set of parameters could be used to fit the conductivity at varying concentrations. The obtained fitted charge is generally different from the structural charge of the nanoparticle, because some ions are condensed on the surface of the nanoparticle and move with a similar velocity; a review of effective charge in colloids may be found in Ref. \cite{Belloni98}. In order to prove the validity of this fit, the conductivity was also computed using a simulation technique that contains much less approximations than the theory. Brownian Dynamics with hydrodynamic interactions was chosen, in the case of a micellar system (TTABr) \cite{DurandVidalJPC06}. Both Fuoss-Onsager theory and Brownian Dynamics relies on the Smoluchowski equation for the motion of the electrolyte and on a hydrodynamic tensor for the description of hydrodynamic interactions \cite{bernard2,ErmakJCP78}. The simulation method had no other approximation, conversely to the theory. For micellar systems, with the use of an effective charge, the comparison worked very well \cite{DurandVidalJPC06}. The order of magnitude of the effective charge was qualitatively reproduced with a Monte Carlo study of ionic condensation. This study was restricted to highly diluted systems. In the case of more concentrated or more asymmetric electrolytes, there is more ambiguity in the choice of the effective charge, and conductivity cannot be fitted with a unique effective charge when the concentration increases \cite{Lucas2014}.

\newpage

There are very important cases for which the Fuoss-Onsager theory has never been validated and for which Brownian Dynamics cannot be used. First, in order to understand the origin of the effective charge and whether condensed ions significantly affect conductivity or not, the direct comparison of the theory with a more accurate simulation method for systems containing highly charged nanoparticles and condensed ions is crucial. Brownian Dynamics is not a reliable tool for such study: The random moves create strong algorithmic instabilities when there is the combination of strong attraction and strong repulsion in a system \cite{MALA}. Secondly, Brownian Dynamics is not adapted to crowded electrolyte solutions, {\em i.e.} highly concentrated electrolyte solutions. This concentration regime is very important for applications, both for industrial  and biological systems. Lastly, testing the validity of the theories in a large range of parameters is also crucial to get more quantitative arguments when specific ion effects are at play. Indeed, the many indications that ion-specific effect beyond ionic size can play a major role in electrolyte behavior will be better understood once the predictions of the primitive models are properly quantified.  
We have worked in our group on the adaptation of a recent mesoscopic simulation methodology, called Multiparticle Collision Dynamics (MPCD), to electrolyte solutions \cite{Batot2013,Zhao2016,Zhao2018}. MPCD does not show any of the limitations of Brownian Dynamics. It is moreover also well suited to confined electrolyte solutions. MPCD allows: (1) to obtain the exact structure of a given model of the electrolyte solution, as would do a Monte Carlo simulation, (2) to account both for hydrodynamic and electrostatic couplings consistently (contrarily to the Fuoss-Onsager theory, it does not separate these effects into independent contributions). Another advantage of MPCD compared to Brownian dynamics is that it does not decompose hydrodynamics interactions in pairwise additive terms: The MPCD algorithm conserves momentum while creating fluctuations, and it is thus equivalent to a Navier-Stokes solver with thermal noise. This makes MPCD valid at high volume fractions and suited to the computation of transport coefficients of solutes\cite{WinklerRev,PaddingPRE06}.  

In this article, we explore electrolytes at the frontiers of the colloidal domain. For systems containing charged nanoparticles, we evaluate the ability of electrolyte theories to describe their dynamic properties. By electrolyte theories, we mean Fuoss-Onsager transport theory combined with Ornstein-Zernike integral equations, with MSA or HNC closures. Our validation procedure is based on a comparison with simulations. We use Multi-particle Collision Dynamics simulations as a benchmark. We first compare the simulation with the theory in conditions for which the theory has been rigorously compared with experimental results, and is known to be valid. For a particular case of charged nanoparticles, the tungstosilicate ion, that is a polyoxometalate ion, we compare the validated theory with recent experimental data. Finally, we discuss a singular transport behavior in suspensions of highly charged but small nanoparticles, where counterions diffuse faster as the nanoparticle concentration increases.  

\section{Methods}
\subsection{Fuoss-Onsager transport theory of the electrical conductivity}
\subsubsection{Framework of the Fuoss-Onsager theory}
The electrical conductivity is a collective transport coefficient of charged systems. It is defined as the ratio of the total charge flux over the applied electric field. In order to compute the charge flux in electrolyte solutions, Fuoss and Onsager assumed that the velocities of ions are overdamped so that  fluctuations of their velocities do not need to be explicitly included\cite{onsa32,onsa45,onsa57}. Therefore, ions of the same type $\alpha$ move with the same velocity $\bf{v}_\alpha$. In order to derive the value of  $\bf{v}_\alpha$, Newton's first law is expressed:
\begin{equation}
m_\alpha\frac{\partial \bf{v}_\alpha}{\partial t} = e_\alpha {\bf E} -\xi_\alpha ({\bf v}_\alpha - {\bf v}_{water}) +{\bf F}_\alpha^{hyd} + {\bf F}_\alpha^{rel} \label{tri1} 
\end{equation}
 
where $m_\alpha$ is the mass of ion $\alpha$, $e_\alpha=z_\alpha e$ with $e$ the elementary charge is its charge. $\xi_\alpha$ is the individual friction on the ion $\alpha$, $\xi_\alpha=k_BT/D_\alpha^{\circ}$, where $D_\alpha^{\circ}$ is the self-diffusion coefficient at infinite dilution of ion $\alpha$, $T$ the temperature and $k_B$ the Boltzmann constant. The Fuoss-Onsager theory introduces four types of forces acting on ions.
The first two forces,$e_\alpha {\bf E}$ and $-\xi_\alpha ({\bf v}_\alpha - {\bf v}_{water})$, are one-body forces on individual ions moving under the presence of an electric field ${\bf E}$, and slowed down by a dissipative friction force due to the solvent. These two forces determine the ideal conductivity, {\em i.e.} the conductivity at infinite dilution, without any interactions between ions. As mentioned in the introduction, interactions between ions play a role even at very low concentration, and correcting forces, namely ${\bf F}_\alpha^{hyd}$ and  ${\bf F}_\alpha^{rel}$, need to be included to account for the deviations from the ideal behavior.
${\bf F}_\alpha^{hyd}$ is the effective force arising from hydrodynamic interactions with all the other ions, 
and ${\bf F}_\alpha^{rel}$ is the effective force arising from direct interactions between ions (such as electrostatic interactions). 
After a characteristic relaxation time scale, usually called the inertial time $\tau_{inertial}=m_\alpha D^{\circ}_\alpha/k_BT$, the velocities of ions in the reference frame of the solvent reach stationary values and become
\begin{equation}
{\bf v}_\alpha = \frac{D_\alpha^{\circ}}{k_BT} \left( e_\alpha{\bf E} + {\bf F}_\alpha^{rel} \right) + \delta {\bf v}_\alpha^{hyd}  \label{tri2} 
\end{equation}
where $\delta {\bf v}_\alpha^{hyd}=
\left(D_\alpha^{\circ}/k_BT\right){\bf F}_\alpha^{hyd}$. 

The electrical conductivity $\chi$ of the solution depends on the ionic velocities, on the ionic concentrations and on the ionic charges through the Ohm law. 
The determination of $\delta {\bf v}_\alpha^{hyd}$  and ${\bf F}_\alpha^{rel}$ is thus the key step to compute the electrical conductivity. 

\subsubsection{Hydrodynamic correction to the ionic velocity}

The hydrodynamic velocity correction  $\delta {\bf v}_\alpha^{hyd}$ can be understood through an analogy with electroosmosis. This term is often called the electrophoretic correction. Around a central ion, the ionic environment is globally charged, with a charge opposite to that of the central ion. The size of the charged volume around an ion is of the order of the Debye length. Within this volume, in the presence of an electric field, the electric forces on the fluid leads to an electroosmotic flow in the frame of the individual ion. This flow imposes a dragging force on the central ion. This is the origin of hydrodynamic corrections to the electrical conductivity. This electroosmotic flow can be approximated by a sum of pairwise hydrodynamic interactions, themselves approximated using the Oseen tensor \cite{ErmakJCP78} between charged points: 

\begin{equation}
    \bf{O}(\bf{r})=\frac{1}{8\pi\eta r}(\bf{1} + \frac{\bf{r} \otimes \bf{r}}{r^2}),
    \label{serge1}
\end{equation}
where $\eta$ is the solvent dynamic viscosity, $\bf{1}$ is the unit tensor. $\bf{r}$ is a vector between a point where an electric force is applied to the fluid and another point where the fluid velocity is computed, $r$ being the distance between these two points. 

To compute $\delta {\bf v}_\alpha^{hyd}$ using the Oseen tensor, a mean field treatment is chosen. The equilibrium $2$-body distributions are used to describe the force distribution in the fluid. As a first approximation, the force at each point is taken equal to the external force $e_\beta {\bf E}$ on the ions due to the electric field \textbf{E} times the ion density: 
\begin{equation}
\delta {\bf v}_\alpha^{hyd}=-\sum_\beta n_\beta \int_{\bf{V}} h_{\alpha \beta}^{\circ}(r) {\bf{O}(\bf{r})} e_\beta {\bf E} \;  \mathrm{d}\bf{r}  \end{equation}
where $\bf V$ is the volume of the solution,  $h_{\alpha \beta}^{\circ}(r)$ is the total pair distribution function between ions of types $\alpha$ and $\beta$ at equilibrium, and $n_{\alpha}$ is the concentration of $\alpha$ ions.
 Then, by performing the angular integration  we get
\begin{equation}
\delta {\bf v}_\alpha^{hyd}=\sum_\beta n_\beta H_{\alpha \beta} e_\beta {\bf E}   \label{hydro_int} 
\end{equation}
with
\begin{equation}
H_{\alpha \beta}=  \frac{2}{3\eta} \int_0^{\infty}rh_{\alpha \beta}^{\circ}(r) \; \mathrm{d}r,
\end{equation}
This restriction on the forces is only valid when the solution is diluted. For concentrated solutions the  interactions between ions must also be taken into account to describe the forces acting on the ions. To this end, we added the forces $ {\bf F}_\beta^{rel} $ to the external forces $ e_\beta \mathbf{E} $ in the eq. 
(\ref{hydro_int}) for $ \delta {\bf v}_\alpha^{hyd} $, through a numerical iterative procedure (see Ref. \cite{Pusset2015,Gourdin15} for more details).  

Interestingly, when $ {\bf F}_\beta^{rel}$ is ignored, the hydrodynamic effect within this theory only depends on the equilibrium distribution $h_{\alpha \beta}^{\circ}(r)$. In that sense, it is similar to an effective or mean force. Therefore, the next important element of the theory is a set of equations to get the equilibrium structure. For the primitive model of electrolyte solutions, integral equations are well suited, and will be described in a subsequent part of the paper. Equilibrium structures computed by simulations can also be used, although for systems with long range correlations, the size of the simulation box may not be sufficient to compute the integral of equation (\ref{hydro_int}). 

\subsubsection{Effective force arising from direct interactions between ions}

At equilibrium, the ionic atmosphere around a given ion is spherically symmetric. This symmetry brakes when an electric field is applied, because of the constant flux of ions in the frame of the central ion. This in turn creates an effective force, which is a dynamical effective force. This effective force is often assumed to result from electrostatic forces, and is called the electrostatic relaxation force. It slows down the motion of ions, and thus decreases the conductivity. We use Onsager methodology to evaluate this force \cite{onsa32}. In this framework, the perturbations of the total correlation functions and of the internal electric field are computed. The total correlation functions $h^{\circ}_{\alpha \beta}$, defined at equilibrium, become the anisotropic functions $h_{\alpha \beta}({\bf r})$ ($=h^{\circ}_{\alpha \beta}(r) + h^{\prime}_{\alpha \beta}({\bf r}) $). The dynamics of $h_{\alpha \beta}^{\prime}({\bf r})$ is described theoretically by a diffusion equation, analogous to the Smoluchowski diffusion equation\cite{Zwanzig69}. Poisson equation couples the perturbation  of the total correlation function $h^{\prime}_{\alpha \beta}$ to the perturbation of the electric field. These functions $h^{\prime}_{\alpha \beta}({\bf r})$ depend on the individual mobilities of ions, and are therefore 
non-equilibrium quantities. 

For a binary electrolyte, i.e. an electrolyte made of one cation $\alpha$ and one anion $\beta$, the functions $h^{\prime}_{\alpha \beta}$ are in the dilute range proportional to $ (D^\circ_\alpha e_\alpha  -
 D^\circ_\beta e_\beta ) / k_B T $ \cite{onsa32}. In comparison with the eq. (\ref{tri2}), we note that $ (D^\circ_\alpha e_\alpha {\bf E} -
 D^\circ_\beta e_\beta {\bf E}) / k_B T $ is actually the speed difference $ {\bf v}_\alpha -{\bf v}_\beta $ when the ions $ \alpha $ and $ \beta$ can be considered ideal ({\em i.e.} when the terms $ {\bf F}^{rel} $ and $ \delta {\bf v}^{hyd}$ can be neglected). When the solutions are concentrated and when they contain highly charged ions, these terms become  important. Consequently, in order to better evaluate the electrostatic relaxation forces, it is necessary to
consider that the functions $h^{\prime}_{\alpha \beta}$ are proportional to the velocity differences $ {\bf v}_\alpha - {\bf v}_\beta $, taking into account the relaxation and hydrodynamic corrections as in eq. (\ref{tri2}). Therefore, the electrostatic relaxation force $\mathbf{F}^{rel}$ depends on the equilibrium 2-body distributions, such as the hydrodynamic correction. It also scales with $1/\eta$ (as far as individual mobilities scale like this, where Stokes friction dominates over other friction mechanisms, such as Enskog friction). 
This approach was first introduced to describe the conductivity of associated salts and of dilute non-aqueous solutions \cite{Ebeling78,Justice78}. Recently, we have used this formalism to describe, from the MSA theory, the electroacoustic signal \cite{Gourdin15}, called the Ionic Vibration Potential. 


Once the non equilibrium total correlation functions $h_{\alpha \beta}^{\prime}({\bf r})$ are determined, the electrostatic relaxation force reads: 

\begin{equation}
\mathbf { F } _ { \alpha } ^ { r e l } = \sum_{ \beta } \int_{ \mathbf{ V } } - \nabla V_{\alpha \beta}(r) n _ { \beta } h _ { \alpha \beta} ^ { \prime } ( \mathbf{ r } ) \mathrm{ d }  \mathbf{ r }
\label{eq7}
\end{equation}
In the case of the conductivity of a binary electrolyte, the relaxation forces of the two types of ions are equals: $\mathbf{F}_{\alpha}^{r e l} =  \mathbf{F}_{\beta}^{r e l}$. 
 This simplification allows to obtain an explicit expression of these forces in terms of integrals dependent  on the equilibrium correlation functions $ h^o_{\alpha\beta}(r) $. This expression is 
\begin{equation}
\mathbf{ F}_{1}^{r e l} = -\frac{K}{1+K} 
\end{equation}
where
\begin{equation}
K  =  {\cal I}\frac{n_\alpha e_\alpha u_\alpha^{\star} +n_\beta e_\beta u_\beta^{\star}}{3\epsilon_0 \epsilon_r \left(D_\alpha^\circ +D_\beta^\circ\right)}   \int_{\sigma}^{\infty} r \; h_{\alpha\beta}^\circ(r) \exp{(-\kappa_q r)}   \; \mathrm{d}r
\end{equation}

with
\begin{equation}
u_\alpha^{\star}  =  e_\alpha\frac{D_\alpha^\circ}{k_B T} + n_\alpha e_\alpha  H_{\alpha\alpha} + n_\beta e_\beta  H_{\alpha\beta}
\end{equation}

and
\begin{equation}
{\cal I}  =  i_0(\kappa_q \sigma) - \frac{4 \pi \epsilon_0 \epsilon_r k_B T}{e_\alpha e_\beta} \kappa_q \sigma^2 i_1(\kappa_q \sigma) 
\end{equation}
with
\begin{eqnarray}
\nonumber
i_0(\kappa_q \sigma) & = & \frac{\sinh{(\kappa_q \sigma)}}{\kappa_q \sigma} \\
i_1(\kappa_q \sigma) & = & \frac{\cosh{(\kappa_q \sigma)}}{\kappa_q \sigma} - \frac{\sinh{(\kappa_q \sigma)}}{\left(\kappa_q \sigma\right)^2}
\end{eqnarray}
and
\begin{equation}
\kappa_q^2 = \frac{1}{\epsilon_0 \epsilon_r k_B T} \frac{n_\alpha e_\alpha^2 D_\alpha^\circ + n_\beta e_\beta^2 D_\beta^\circ}{D_\alpha^\circ + D_\beta^\circ}
\end{equation}
where $\sigma$ is the minimum distance of approach between ions of types $\alpha$ and $\beta$, $\varepsilon_0$ is the permittivity of vacuum, $\varepsilon_r$ the relative permittivity if the solvent (here pure water).

When the electrolyte solution contains more than two types of ions, $ {\bf F}_\alpha^{rel} $ is computed using an iterative approach \cite{Pusset2015}.

\subsection{Determination of the equilibrium correlation functions from integral equations}
\subsubsection{Primitive model of electrolyte solutions}
The equilibrium structure of the electrolyte solution depends both on the model and on the theory used to "solve" it. Within the so-called "primitive model" (PM) of electrolyte solutions, the molecular nature of the solvent is ignored, and the latter is replaced by a continuum solely characterized by its dielectric permittivity $\varepsilon=\varepsilon_0\varepsilon_r$. In the present article, we study
\begin{itemize}
\item[(i)] $1-1$ electrolyte solutions through the restricted primitive model, with microions of same size (diameter $\sigma$), same charge  ($e_{+/-}={+/-}1e$),
number of ions $N_{+/-}=N$), and in a
volume $V$, such that the number densities are $n_{+/-}=(N_{+/-})/V$,
\item[(ii)] dispersions of  monodisperse positively charged 
nanoparticles (typically micelles or globular proteins) modelled as a system of
$N$ hard spheres of diameter $\sigma_{NP}$ and charge $e_+=Ze$, surrounded by smaller monovalent counterions ($e_{-}=-1e$). Overall charge neutrality of the dispersion implies $Zn_+ -n_- = 0$. 
\end{itemize}
The pair interaction potential between ions of types $\alpha$ and $\beta$ is:
\begin{equation}
\label{PM}
\left\{
\begin{array}{ll}
{V}_{\alpha \beta}(r) = \frac{e_{\alpha}e_{\beta}}{4\pi\varepsilon_0\varepsilon_r}
\frac{1}{r}  & \;\;{\rm for }\;\; r > \frac{1}{2}(\sigma_\alpha+\sigma_\beta) \\
{V}_{\alpha \beta}(r) = \infty &
\;\;{\rm  for  } \;\;r < \frac{1}{2}(\sigma_\alpha+\sigma_\beta)\\
\end{array}
\right.
\end{equation}
In what follows, the characteristic electrostatic length scale, the Bjerrum length $l_b={e^2}/{(4\pi\varepsilon_0\varepsilon_r k_BT)}$,  is equal to $0.71$ nm, 
corresponding to water at room temperature.

\subsubsection{Integral equation approach}
It has been proven that pair distribution functions are unambiguously related to a model (or a set of effective pair potentials between particles) \cite{Henderson74,Chayes84}. In practice, the structure/potential relationship is based on fluid integral equations for the pair structure, 
which are based on the Ornstein-Zernike (OZ) equation, and approximate closure relations\cite{hansen}.
Analytical expressions of the total pair distribution function between ions of types $\alpha$ and $\beta$, $h^\circ_{\alpha \beta}(r)=g_{\alpha \beta}(r)-1$, can be derived by combining the OZ equation with the Mean Spherical Approximation (MSA). The OZ equation reads
\begin{equation}
    h_{\alpha \beta}^{\circ}(r) = c_{\alpha \beta}^{\circ}(r) + \sum_\gamma n_\gamma \int_{\bf V} c_{\alpha\gamma}^{\circ}( s)h_{\gamma\beta}^{\circ}(\vert{\bf r}-{\bf s}\vert) \; \mathrm{d}{\bf s} 
    \label{val2}
\end{equation}
where $c_{\alpha \beta}({\bf r})$ is the direct correlation function between $\alpha$ and $\beta$. For distances higher than the minimal distance of approach between ions, the direct correlation function reads in the case of the primitive model in the MSA:
\begin{equation}
 c_{\alpha \beta}^{\circ}=-\frac{V_{\alpha \beta}}{k_BT}
\end{equation}
For distances smaller than the minimal distance of approach, the closure $h_{\alpha \beta}^{\circ}=-1$ is assumed. 

Within the MSA, the integrals used to compute the electrical conductivity of the solutions from the Fuoss-Onsager theory described in the previous section can be derived semi-analytically. Then, explicit expressions of the hydrodynamic integrals,  necessary to compute $\delta {\bf v}_\alpha^{hyd}$ with eq. (\ref{hydro_int}), have been given previously (eq. (5) of Durand-Vidal {\em et al}\cite{DurandVidalJPC06}).  In the same way, semi-analytic expressions of the integrals involved in the evaluation of the relaxation forces can be deduced. 



The MSA  is a linearised theory that is particularly suitable when the interaction potential is small compared to  $ k_B T$. However, it is known that MSA does not lead to quantitative predictions  of the pair correlation functions
for highly charged particles. To determine if the charge is low enough for the MSA to be valid, a simple condensation criterion can be used: we should have  $Zl_b/R<1$, where $l_b$ is the Bjerrum length\cite{Belloni98}. For instance, a particle of radius $R=1$~nm and charge $Z=20$ in water at room temperature with $l_b\simeq 0.7$~nm is considered as highly charged. 
In this study, we thus also used a numerical estimation of the correlation functions, based on the OZ equation, 
with the HyperNetted Chain (HNC) equation as a closure\cite{HNC}. This more accurate level of theory is expected to be appropriate for Coulombic systems, even at relatively high electrostatic coupling. 
The HNC closure reads:
\begin{equation}
g_{\rm \alpha \beta}(r)=\exp[-V_{\rm \alpha \beta}(r)/k_{\rm B}T + h^{\circ}_{\rm \alpha \beta}(r) -c_{\rm \alpha \beta}^{\circ}(r)]
\label{hnceq}
\end{equation}
At high packing fractions of charged nanoparticles,  where hard core correlations between nanoparticles become predominant, the HNC closure usually fails to describe faithfully the correlation functions and 
 may be improved by including an estimate of the
\emph{bridge} function \cite{Rosenfeld79}.  However, for all the systems  investigated here, we could not detect any significant differences between the structure predicted by HNC and that deduced from the numerical simulations. Therefore we did not use any bridge function. 

Nevertheless, 
solving OZ and HNC numerically can be a non trivial task. Several numerical instabilities appear, in particular at low electrolyte concentration, when screening is less important and interactions are at longer range \cite{Rasaiah72}. Divergence issues can be avoided by increasing progressively the charge of the species. For a few systems, we could not get results with this method, and therefore we used the distribution functions from the simulations as input of the Fuoss-Onsager transport theory. 

In summary, our theoretical approach deals with two components. First, the equilibrium correlation functions are obtained from the OZ equations,
solved with the MSA or HNC closure relation. These correlations are used within a Fuoss-Onsager (FO) treatment of hydrodynamic and electrostatic corrections to the ideal electrical conductivity. Moreover, in this study, we take into account, in a self-consistent way, deviations from the ideal behavior to evaluate the relaxation force.
To distinguish this application of the  FO theory from the one  previously used \cite{bernard2}, we call this new approach Self-consistent Fuoss-Onsager (SFO). Depending on the choice of the closure relation, we will refer in what follows either to MSA-SFO, or to HNC-SFO. The parameters required to compute the electrical conductivity from MSA-SFO or HNC-SFO are the ionic concentrations, the self-diffusion coefficients of ions at infinite dilution, the diameters and the charges of ions.

\section{Multiple Particle Collision Dynamics}
Multiparticle Collision Dynamics, MPCD,  is the most adapted technique to test the validity of the approximations of the aforementioned theories (MSA-SFO or HNC-SFO theories) as it accounts both for hydrodynamic couplings between ions and for electrostatic interactions. It is an explicit solvent method, although the solvent is the simplest one: it is a low density fluid where momentum is transported through ballistic motions and collisions between fluid particles \cite{WinklerRev,PaddingPRE06}. This fluid can be coupled to solute particles in various ways. For asymmetric electrolytes, we recently proposed to couple the dynamics of the charged nanoparticles to the fluid using an explicit  fluid/nanoparticle interface on which the fluid is reflected, while small ions are described very similarly to MPCD fluid particles \cite{Zhao2016}. 

	\subsection{MPCD algorithms }
	\subsubsection{Case of a pure MPCD fluid}

The fluid in MPCD is represented by point-like particles. Their positions and velocities evolve in two steps \cite{PaddingPRE06}. 
In the  {\em streaming step}, positions and velocities of each fluid particle $i$ are propagated by integrating Newton's equations of motion :
	\begin{equation}
		\mathbf{r}_{i}(t + \delta{}t_{c}) = \mathbf{r}_{i}(t) + \mathbf{v}_{i}(t)\delta{}t_{c}
		\label{streamfluid}
	\end{equation}
where $\mathbf{r}_{i}$, $\mathbf{v}_{i}$ are respectively the position and the velocity of particle $i$; and $\delta{}t_{c}$ is the time step. 
A second step, the  {\em collision step}, enables local momentum exchanges between the fluid particles. The simulation box is partitioned into cubic cells, called collision cells. A randomly oriented axis is defined for each collision cell, and the velocities of fluid particles relative to the velocity of the center of mass of the cell are rotated by an angle $\alpha$ around this axis:
	\begin{equation}
		\mathbf{v}_{i}(t + \delta{}t_{c}) = \mathbf{v}^{cell}_{c.o.m}(t) + \mathcal{R}_{\alpha}[\mathbf{v}_{i}(t)-\mathbf{v}^{cell}_{c.o.m}(t)]
		\label{collisionRule}
	\end{equation}
where $\mathcal{R}_{\alpha}$ is the rotation matrix and $\mathbf{v}^{cell}_{c.o.m}$ the velocity of the center of mass of the cell. The angle $\alpha$ is a fixed parameter. A random shift of the collision grid is performed at each collision step to ensure galilean invariance\cite{WinklerRev,GalInvarianceSRD}.

Analytical formulas for the viscosity and transport coefficients of the MPCD fluid can be derived. They depend on the number of solvent particles per cell $\gamma$, on the rotation angle $\alpha$, and on the time step $\delta t_c$\cite{PaddingPRE06,WinklerRev}.
It is convenient to use the fluid particle mass ${m_f}$ as the mass unit, the size of the collision cells $a_0$ as the length unit, and $k_BT$ as the energy unit. The time unit is then
\begin{equation}
t_0 = a_0\sqrt{\dfrac{m_f}{k_BT}}.
\end{equation} 

\subsubsection{Embedded particles in the MPCD bath: Collisional Coupling }

Within the {\em collisional coupling} scheme, solute particles interact with each other through a classical force field, and participate to the collision step with solvent particles. They have a mass larger than fluid particles. In the collision step, the velocities of all solvent and solute particles in each collision cell are thus updated following Eq. (\ref{collisionRule}).
In the streaming step, the positions of solvent fluid particles are updated following Eq. (\ref{streamfluid}), whereas the position ${\bf R}_j$ and velocity ${\bf V}_j$  of solute $j$ are propagated with the velocity Verlet algorithm often used in standard Molecular Dynamics (MD) simulations.  The force between solute particles is derived from their interaction potential. Details about this simulation scheme can be found in several reviews \cite{RipollPRE05,WinklerRev}. 


The clear advantage of this coupling method is that it is very efficient from the computational point of view. One drawback is that, as we showed in a recent article \cite{Zhao2018}, the hydrodynamic radius $a_{hyd}$ of solute particles is almost constant at the scale of the MPCD collision cell size a$_0$, 
of the order of $0.3$a$_0$. 
In the case where the hydrodynamic radius of the real solute is known, this constraint imposes the size of the MPCD 
cell in physical units, for instance in nanometers. 
For a given structural model of the solute (for instance the radius for a hard sphere model of solutes, or the size parameter of a purely repulsive short ranged interaction potential), 
the resolution of the MPCD cells relative to the size of the particles is then imposed. 
Also, the relatively small value of the hydrodynamic radius compared to the collision cell might lead to artefacts
if the hydrodynamic radius is equal to the structural radius. In this case, 
two or more solute particles can be located in the same MPCD collision cell at the same time, which yields to an underestimation of the diffusion coefficient of the solute. In the case where the structural diameter of ions is the size parameter $\sigma_{WCA}$ of a purely repulsive Week-Chandler-Andersen (WCA) interaction potential, 
we have shown in our previous study \cite{Zhao2018} that the best set of parameters correspond to $\sigma_{WCA} = 1.5$a$_0$.  
This set of constraints ($\sigma_{WCA} = 1.5$ a$_0$ and $a_{hyd} = 0.3$ a$_0$) is not a technical problem for the current study, as it is straightforward to change independently the values of the structural and of the hydrodynamic radii within the theories. 

\subsubsection{Stochastic Reflection rules}

Another family of coupling methods between the MPCD fluid and solutes intends to reproduce no-slip boundary conditions at the surface of the solute (or of a wall). This condition is a better representation of solutes as colloids or nanoparticles, since attractive short-ranged interactions with the solvent
are expected to stick solvent molecules at the surface of the particles~\cite{PaddingJPCM05,WhitmerJPCM10} in most real systems. In the present study, we used the Stochastic Reflection Rules algorithm (SRR). The SRR for solvent particles around nanoparticles was first proposed by Inoue {\em et al} \cite{Inoue2002}. It was later refined by Padding and Louis \cite{PaddingJPCM05} (we use here this latter version of the algorithm). The reader is referred to these publications for more details. Briefly, within this scheme, when a solvent particle enters a solute particle,
the time and position of the impact is computed, the solvent particle is restored to this impact point and is given a random velocity obtained through a half-plane Maxwell-Boltzmann distribution. The velocity of the solute evolves to conserve linear and angular momentum. 
Within this methodology, it is not necessary to divide the streaming step into smaller MD steps, except if the density of solutes is important. 


In the case of systems containing charged nanoparticles, counterions are attracted by the nanoparticle, so that they are concentrated close to the surface of the nanoparticles.
Using a repulsive interaction potential between small ions and nanoparticles to mimic the excluded volume of the nanoparticle renders the description of the dynamics of electrostatically condensed ions challenging. In order to address this issue, we proposed in a recent article to apply Stochastic Reflection Rules to counterions in the vicinity of nanoparticles\cite{Zhao2016}. The same algorithm is thus  used for counterions and solvent particles when they encounter a nanoparticle. This method compared very well with the more expensive use of explicit forces, and is used in the present study.

\subsection{Summary of the numerical simulation methodology}

We study in this paper two families of systems: the first one contains only the small ions of a simple 1-1 electrolyte solution, and the second one contains charged nanoparticles of different charges and sizes, together with their counterions, described as small monovalent ions. In what follows, both ions of the 1-1 electrolyte solutions are described within the collision coupling rule. 
Where charged nanoparticles are under study, we used a mixed coupling methodology introduced in our recent article \cite{Zhao2016}. In this case, counterions are coupled to the solvent within the collision steps, {\it i.e} we use the CC algorithm for these small solutes. The nanoparticles are coupled to the solvent bath by using
the Stochastic Reflection Rules mimicking no-slip boundary conditions (SRR algorithm). Counterions also interact with the nanoparticle through the SRR scheme. In all cases, periodic boundary conditions were
applied to the cubic simulation box of volume $V=L_{box}^3$, and an Ewald summation
was used to compute the Coulombic interactions of the infinite array of
periodic replicae of the simulation box.
We had in previous works computed the values of the self-diffusion coefficients at infinite dilution of solutes described either with the collision coupling scheme or the SRR coupling algorithms\cite{Zhao2016,Zhao2018}. These values are denoted by $D_\alpha^\circ$ in what follows and are used to compute the ideal electric conductivity $\chi^\circ$.

\subsection{Parameters of MPCD simulations}	

The parameters related to the solvent are chosen to reproduce hydrodynamic interactions typical of a liquid. 
As shown by Ripoll et al. \cite{RipollPRE05}, simulating a liquid-like fluid while minimizing the computational cost leads to the following set of parameters: $\{\alpha=130^{\circ},\gamma=5,\delta t_c=0.1t_0\}$. For this choice of parameters, the kinematic viscosity of the  fluid is $\nu=0.809 \;a_0^{2}~t_0^{-1}$, so that the dynamic viscosity $\eta$ is equal to $4.045\;m_f~a_0^{-1}~t_0^{-1}$.
The solutes coupled through the MPCD fluid during the collision steps (CC) are dynamically characterized through their mass. We chose here a mass $M=10m_f$, which is twice the average mass of solvent fluid particles within a collision cell. For more discussions on this parameter, read Ripoll et al. \cite{RipollPRE05}. 
In order to avoid divergence of the energy of the simulation box, the MD step $\delta{}t_{MD}$ used to integrate the equation of motion for solutes was in some cases smaller than the streaming time step $\delta{}t_{c}$ between two collisions. Here, $\delta{}t_{MD}$ was empirically chosen based on the stability of the total energy of the system, from $0.1 \delta{}t_{c}$ to $1 \delta{}t_{c}$ depending on the system. No thermostat was used. 

With these parameters for the MPCD fluid, embedded particles of mass $M=10 \;m_f$ who are coupled with the fluid within the collision step have an infinite dilution diffusion coefficient $D_\alpha^\circ$ of $4.175 10^{-2} \;a_0^{2}~t_0^{-1}$\cite{Zhao2018}. The diameter of SRR particles characterizing the stochastic reflection of the solvent is $\sigma_{NP}=4 \;a_0$, and their mass is $M=150 \;m_f $, which corresponds to an infinite dilution diffusion coefficient $D_\alpha^\circ$ of $1.09 10^{-2} \;a_0^{2}~t_0^{-1}$ \cite{Zhao2016}.

\subsection{Computation of the electrical conductivity and related transport coefficients from simulation trajectories}

Transport coefficients of solutes are computed from equilibrium simulations, {\em i.e.} without any applied electric field, using a Kubo formula based on the autocorrelation function of the electrical current. The electrical conductivity of the system reads:
\begin{equation}
	\chi = \frac{1}{3k_BTV} \int_0^{\infty} {\rm d}t \, 
			\langle \sum_{i=1}^N e_i{\mathbf V}_i(t_0) \cdot \sum_{j=1}^N e_j{\mathbf V}_j(t_0+t)
			\rangle_{t_0},  
\end{equation}
where ${\mathbf V}_i(t)$ is the velocity of solute particle $i$ at time $t$, and $N$ the total number of solute particles in the simulation box. 
This collective correlation function converges slowly, and is noisier than individual transport functions, such as the individual velocity autocorrelation function computed to get the diffusion coefficients. 
In principle, it depends on the size of the simulation box, similarly to the diffusion coefficient \cite{Yeh2004}. In the present study, we chose large box sizes $L_{box}$, so that the finite box size correction to the diffusion coefficient $2.837 k_BT/(6 \pi \eta L_{box})$ with $\eta$ the viscosity,  is lower than $3\%$ (which means that $2.837 a_{hyd}/L_{box}<0.03$).

Two other transport coefficients have been computed. The fist one is the electrophoretic mobility of species $\alpha$, denoted by $\mu_{\alpha}$. It is an individual transport coefficient related to the cross-correlation of the electric current with the individual charge transport:

\begin{equation}
	\mu_{\alpha} = \frac{1}{3Vk_BT} \int_0^{\infty} {\rm d}t \, 
			\langle \sum_{i=1}^N e_i{\mathbf V}_i(t_0) \cdot e_{\alpha}{\mathbf V}_j(t_0+t)
			\rangle_{t_0, j}.
\end{equation}
The average is made over all particles $j$ of type $\alpha$.

The second one is the Ionic Vibration Potential (IVP): It is a transport coefficient corresponding to the electric potential difference appearing in a solution in response to an applied acoustic field\cite{Pusset2015}. The IVP can be deduced with a good approximation from the electrical conductivity $\chi$ and the individual mobilities $\mu_{\alpha}$, using the following formula:

\begin{equation}
	IVP = \frac{\Delta P}{\chi\rho_s}
	\left(n_{\alpha}  \mu_{\alpha} M_{\alpha}^{eff}
	+	n_{\beta}  \mu_{\beta} M_{\beta}^{eff}\right) 
\end{equation}
where $\Delta P$ is the amplitude of the pressure wave corresponding to the acoustic field, and ${\rho_s}$ is the density of the solution, the effective mass $M_{\alpha}^{eff}$ is the mass of the ions, minus the mass of solvent corresponding to the ionic volume \cite{Hermans38,Pusset2015}. 

\section{Results}
\subsection{Simulation and theory agree for 1-1 electrolytes}

We performed a series of simulations for a simple 1-1 electrolyte in water at ambient temperature, for various concentrations. 
The electrolyte is described by the restricted Primitive Model (see eq. \ref{PM}). 
Both ions have a structural diameter $\sigma_{HS}=0.25$ nm. 
The size of the collision cell is chosen so as to maximize hydrodynamic interactions \cite{Zhao2018}: $a_0=0.66\times\sigma_{HS}=0.167$ nm. In this case, the hydrodynamic radius is thus $a_{hyd}=0.050$~nm\cite{Zhao2018}. 
The box length is kept equal to $32$a$_0=5.3$~nm, so that
the number of ions in the simulation box varies from $56$ to $1188$, when the concentration increases from $0.3$ to $6.3$ mol~dm$^{-3}$. 
For each system, $30$ independent simulations are run for $2\times10^{5}t_0$. The value of $t_0$ in real units may be obtained by mapping the viscosity of the SRD fluid to that of water at room temperature. It corresponds to a physical time of $23$~ns. 

\begin{figure}[h]
\centering
  \includegraphics[scale=0.38]{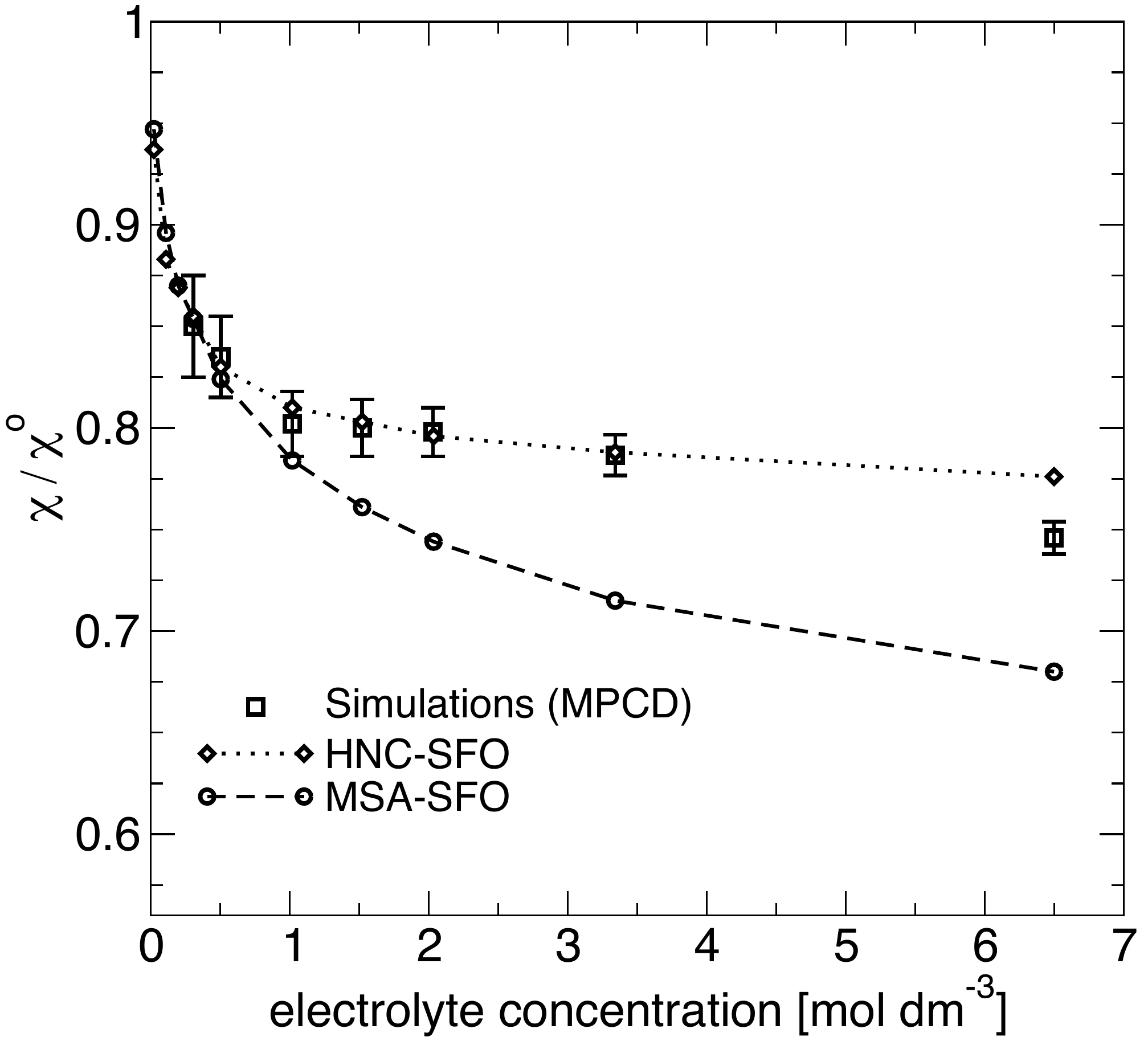}
  \caption{Electrical conductivity $\chi$ of a $1-1$ electrolyte, divided by the value at infinite dilution $\chi ^ { \circ }$, as a function of the electrolyte concentration. Open squares: MPCD simulations (error bars represent the standard deviation over $30$ independent simulation runs); Open circles and diamonds: Self-consistent Fuoss-Onsager (SFO) transport theory with equilibrium correlation functions  deduced from integral equations, using two closure equations: HNC (diamonds), and MSA (circles). 
  }
  \label{figure1}
\end{figure}

Fig.\ref{figure1} displays the electrical conductivity divided by its ideal value as a function of the electrolyte concentration, obtained from MPCD numerical simulations and from the HNC-SFO and the MSA-SFO theories. 
The structural radii and self-diffusion coefficient at infinite dilution used  in the SFO treatments were exactly the same as those of simulations. 
As discussed above, distribution functions obtained from the HNC closure are in principle the same as those obtained by numerical simulation for a given model, contrarily to those obtained by the MSA closure. Indeed, we checked that the distribution functions from  HNC theory and from the simulations coincide. 
In principle, the SFO can thus be combined either with distribution functions from HNC or from the simulations. We have checked for several concentrations that the electrical conductivities computed in both ways agree, but we display the values obtained from HNC in Fig.\ref{figure1}, as the functions obtained in this way do not suffer any noise. As showed on Fig.\ref{figure1}, the electric conductivities computed from HNC and from MPCD agree very well on the whole concentration range. 
 There is only a small deviation of about $4 \%$ for the very high concentration of $6.5$~mol~dm$^{-3}$. 

However, the agreement between MSA-SFO and numerical simulations is good only for concentrations smaller than $1.0$~mol~dm$^{-3}$, as shown in Fig.\ref{figure1}. In the existing literature, MSA-SFO and HNC-SFO were indeed proved to predict faithfully the conductivity of simple $1-1$ electrolytes in this range of concentration\cite{bernard2}. MPCD also predicts the same values in this concentration domain. Our study shows that the limitations of the MSA-SFO come from the limitations of the MSA to describe equilibrium properties, and not from the SFO treatment of electrical transport. Nevertheless, the small discrepancy between HNC-SFO and MPCD at the highest investigated concentration is probably due to the SFO treatment, since the equilibrium distribution functions are the same.

\subsection{HNC-SFO theory quantitatively describes solutions of nanometric ions}
\subsubsection{Case of $1-4$ electrolytes}
We turn to the important case of suspensions of nanoparticles. The size domain explored here corresponds to radii around one nanometer. There exist many synthetic polyoxometalates and other kinds of inorganic nanoparticles in this range, as well as organic materials, including small globular proteins and micellar systems. We first studied small nanoparticles, with similar size and charge as tungstosilicate anions ([SiW$_{12}$O$_{40}$]$^{4-}$). The diameter of such particle is around $1$~nm, and its charge is $-4$e.

We performed a series of simulations for this $1-4$ electrolyte, for various concentrations, using the primitive model (see eq. \ref{PM} for the interaction potential).  
 Nanoparticles have a structural diameter of $\sigma_{NP}=1.05$~nm, while  counterions have a structural diameter of $\sigma_{ci}=0.35$~nm. 
The size of the collision cell is again $a_0=0.77\times\sigma_{ci}=0.270$~nm. In this case, the hydrodynamic radius of counterions in MPCD simulations is roughly $a_{hyd,ci}=0.090$~nm, while the hydrodynamic radius of the nanoparticles is $a_{hyd,NP}=0.345$~nm \cite{Zhao2016,Zhao2018}. We used the same values in the theoretical treatments.
The box length is kept equal to $32$~a$_0=8.64$~nm. 
The number of nanoparticles in the simulation box varies from $4$ to $32$, corresponding to volume fractions of nanoparticles $\phi$ from about $0.5 \%$ to $4 \%$ ($\phi=N\pi\sigma_{NP}^3/6V$). This corresponds to relatively high volume fractions for conductivity experiments in nanoparticle suspensions.
For each system, $30$ independent simulations are run for $2\times10^{5}t_0$. 

\begin{figure}[h]
\centering
  \includegraphics[scale=0.38]{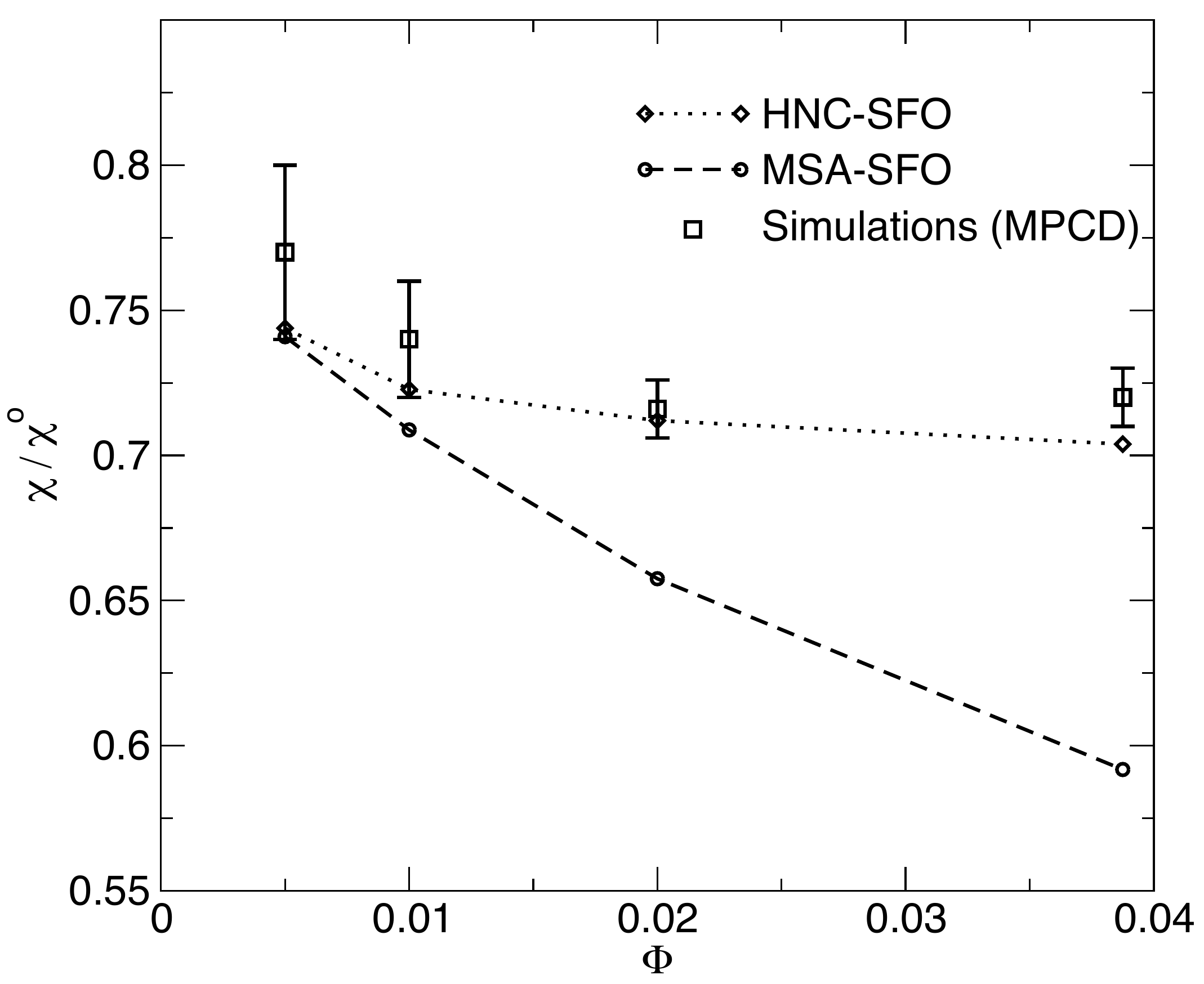}
  \caption{Electrical conductivity $\chi$ of a $1-4$ electrolyte, divided by the value at infinite dilution $\chi ^ { \circ }$, as a function of the volume fraction in nanoparticle $\phi$. Open squares: MPCD simulations; Open diamonds and circles: Self-consistent Fuoss-Onsager transport theory, with HNC closure (diamonds), and MSA closure (circles).}
  \label{figure2}
\end{figure}

In Fig. \ref{figure2}, the electrical conductivity as a function of the volume fraction of nanoparticles is presented. The predictions from the Self-consistent Fuoss-Onsager (SFO) theories are presented in two cases: (i) when correlation functions deduced by MSA are used, and (ii)  when they are deduced by HNC. For volume fractions higher than $0.5 \%$, the SFO theory predicts conductivity values that are significantly smaller with MSA than with HNC. This discrepancy is not surprising as MSA is not expected to work at high concentration. For the largest density, the difference between MSA and HNC results is of about $20 \%$. Nevertheless, the use of MSA integral equations, which yield semi-analytical formulas that are much easier to implement, does not seem to be a limitation up to $\phi=0.5 \%$, an already quite large packing fraction in the context of the analysis of nanoparticle suspensions. As the HNC and MPCD correlation functions are remarkably close, the comparison of HNC-SFO theory and of the simulation in the case of larger packing fractions reveals the limitations of our implementation of the Fuoss-Onsager transport theory. 
For the packing fractions between $0.01$ and $0.04$, the difference between the simulated values and the theoretical values is of about $2 \%$. 

This is the first quantitative comparison of theory and simulation for the conductivity of nanoparticle suspensions at such high packing fraction. We show that for $1-4$ electrolytes, where the nanoparticle is a big tetravalent ion of diameter about one nanometer, the classical electrolyte theories lead to very satisfactory agreement with simulations, up to $0.5 \%$ in packing fraction for MSA-SFO transport theory and up to $4\%$ for the HNC-SFO transport theory.  


One example of $1-4$ electrolyte in water is the aqueous solution of potassium tungstosilicate $\mathrm {K}_{4} \mathrm { SiW } _ { 12 } \mathrm { O } _ { 40 }$.  We propose here to re-interpret experimental results from our group, by Pusset, Dubois and coworkers \cite{Pusset2015} using the HNC-SFO theory. Indeed, a series of measurements of the Ionic Vibration Potential (IVP) was done recently for solutions of tungstosilicate ions up to $\phi=0.06$, and interpreted in ref. \cite{Pusset2015} using the MSA-SFO theory. For the highest volume fractions investigated in this previous work, the use of MSA may be  questionable.

\begin{figure}[h]
\centering
  \includegraphics[scale=0.38]{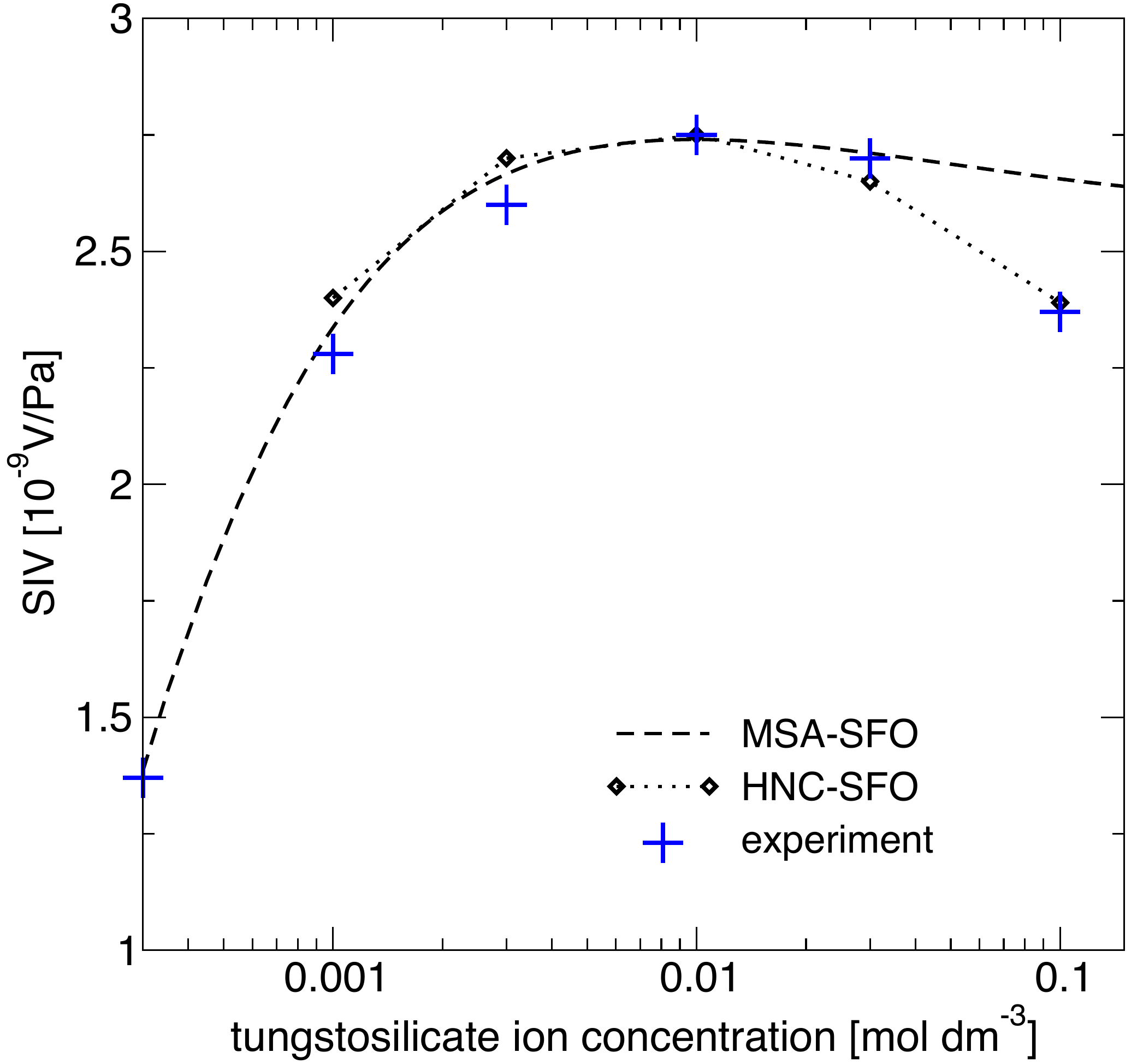}
   \caption{Ion Vibration Signal (S$_{IV}$) of potassium tunsgtosilicate solutions ($\mathrm {K}_{4} \mathrm { SiW } _ { 12 } \mathrm { O } _ { 40 }$). S$_{IV}$ is defined as the ratio of the ion vibration potential over the amplitude of the pressure wave in an electroacoustic experiment.  The detail of the experiments is described in ref.\cite{Pusset2015}. The experimental values are compared with the predictions from the SFO theory, with total equilibrium pair correlation function $h_{ij}^{\circ}(\bf{r})$ computed with HNC (diamonds), or with MSA (dashed line).}
  \label{figure3}
\end{figure}

Fig.\ref{figure3} displays the IVP computed by MSA-SFO and by HNC-SFO compared to experimental values as a function of the concentration in tungstosilicate ions. The experimental results are taken from ref. \cite{Pusset2015}. 
In the experimental system, tungstosilicate anions are surrounded by both $K^+$ and $H^+$ ions. The concentration $[H^+]$ is assumed to be constant, $[H^+]=5\times10^{-3}$~mol~dm$^{-3}$. The parameters of the primitive model of this solution are $Z=-4e$, $\sigma_{NP}=1.30$ nm, and $D_{NP}^\circ=0.56 \times10^{-9}$~m$^2$~s$^{-1}$ for the tungstosilicate anion, $D^\circ=9.311\times10^{-9}$~m$^2$~s$^{-1}$ for the $H^+$ ions and $D^\circ=1.957\times10^{-9}$~m$^2$~s$^{-1}$ for $K^+$. As shown on this figure, experiments, MSA-SFO and HNC-SFO coincide for concentrations lower than $0.03$~mol~dm$^{-3}$. But, for the most concentrated point ($c=0.1$~mol~dm$^{-3}$, i.e. $\phi=0.06$), the use of MSA integral equations is clearly not accurate, while the use of HNC equations yields a result that is remarkably close to the experimental value. This confirms the ability of the theory at the HNC level to exploit experimental data in dense suspensions of small nanoparticles.  

\subsubsection{Case of 1-8 electrolytes}
 There exist  charged polyoxometalate nanoparticles with a charge higher than the tungstosilicate ion, as well as many examples of highly charged nanoparticles for which counterions are electrostatically condensed at the surface of the nanoparticle \cite{Belloni98,DurandVidalJPC06,Lucas2014}. We turn now to this class of systems. We chose to keep the exact same size as in the previous part ($\sigma_{NP}=1.05$ nm, $\sigma_{ci}=0.35$ nm), as well as all parameters related to the fluid. In order to get a clear electrostatic condensation, while staying similar to existing systems, we study the case where the nanoparticle has a charge of $-8$e.

\begin{figure}[h]
\centering
  \includegraphics[scale=0.38]{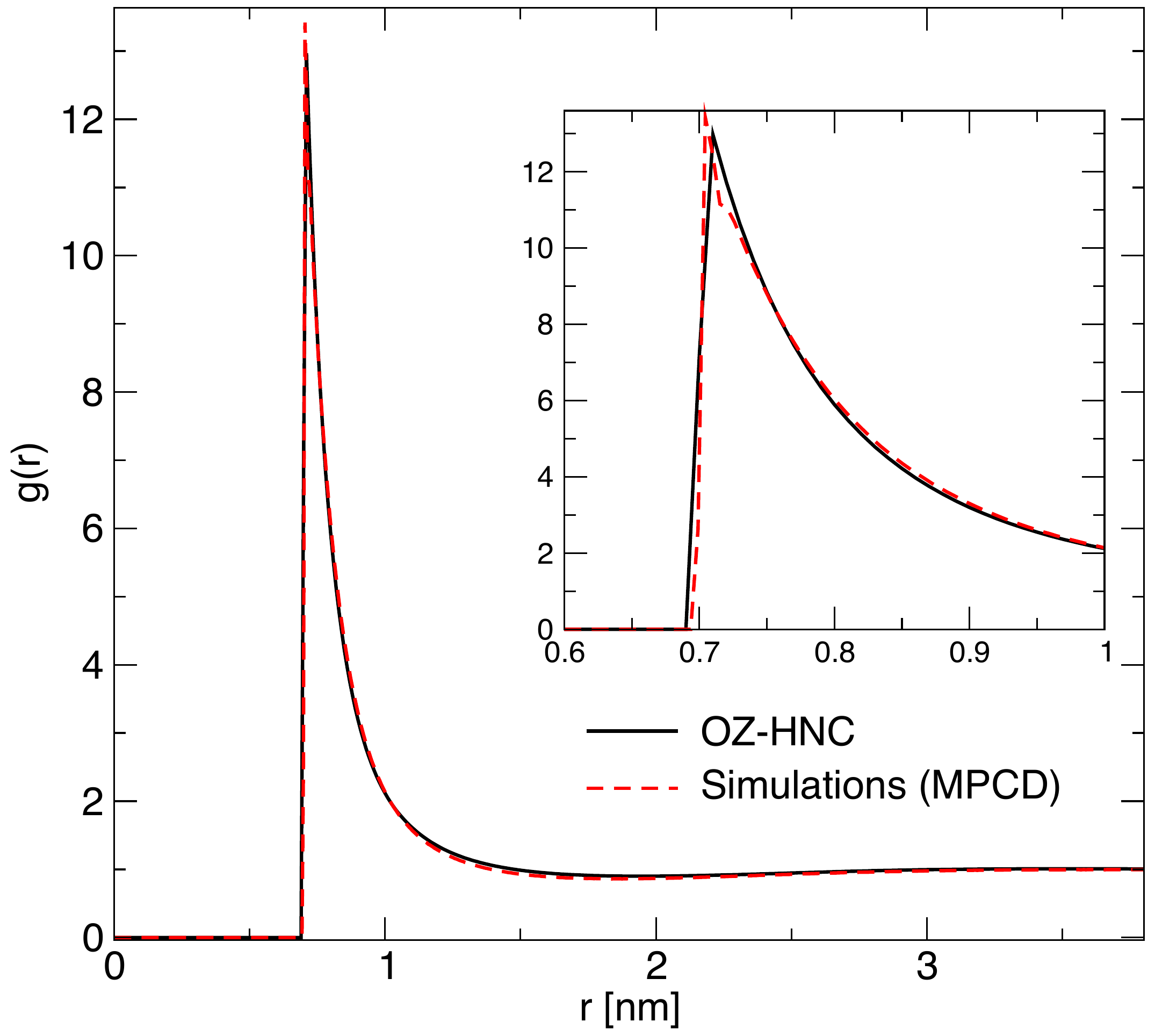}
  \caption{Pair radial distribution function g(r) between the nanoparticles and the counterions, in the case where the nanoparticle is highly charged ($Z_{NP}=8$ for a diameter $\sigma_{NP}=1.05$ nm), for a volume fraction in nanoparticles equal to $0.02$. The black curve corresponds to the results of integral equations (HNC), while the red dashed line corresponds to the results of the simulations (MPCD).}
  \label{figure4}
\end{figure}

For such asymmetric electrolytes, the numerical resolution of HNC integral equations may be technically impossible because the iterative solution of the integral equations does not converge. At large concentrations, as screening increases, this divergence issue disappears. Therefore, we implemented the self-consistent Fuoss-Onsager theory with the static correlation functions obtained from MPCD simulations at low concentrations, and obtained from  HNC at large concentrations ($\phi > 0.02$). We checked that the results were similar for the intermediate concentration case. 
The ability of HNC to predict the correct structural properties of the suspension was tested against numerical simulations. The pair distribution functions $g(r)$ obtained from theory and simulation are shown in Fig. \ref{figure4} for the packing fraction of $0.02$. It appears that the structure predicted by the theory is remarkably close to that of the simulations. 

\begin{figure}[h]
\centering
  \includegraphics[scale=0.38]{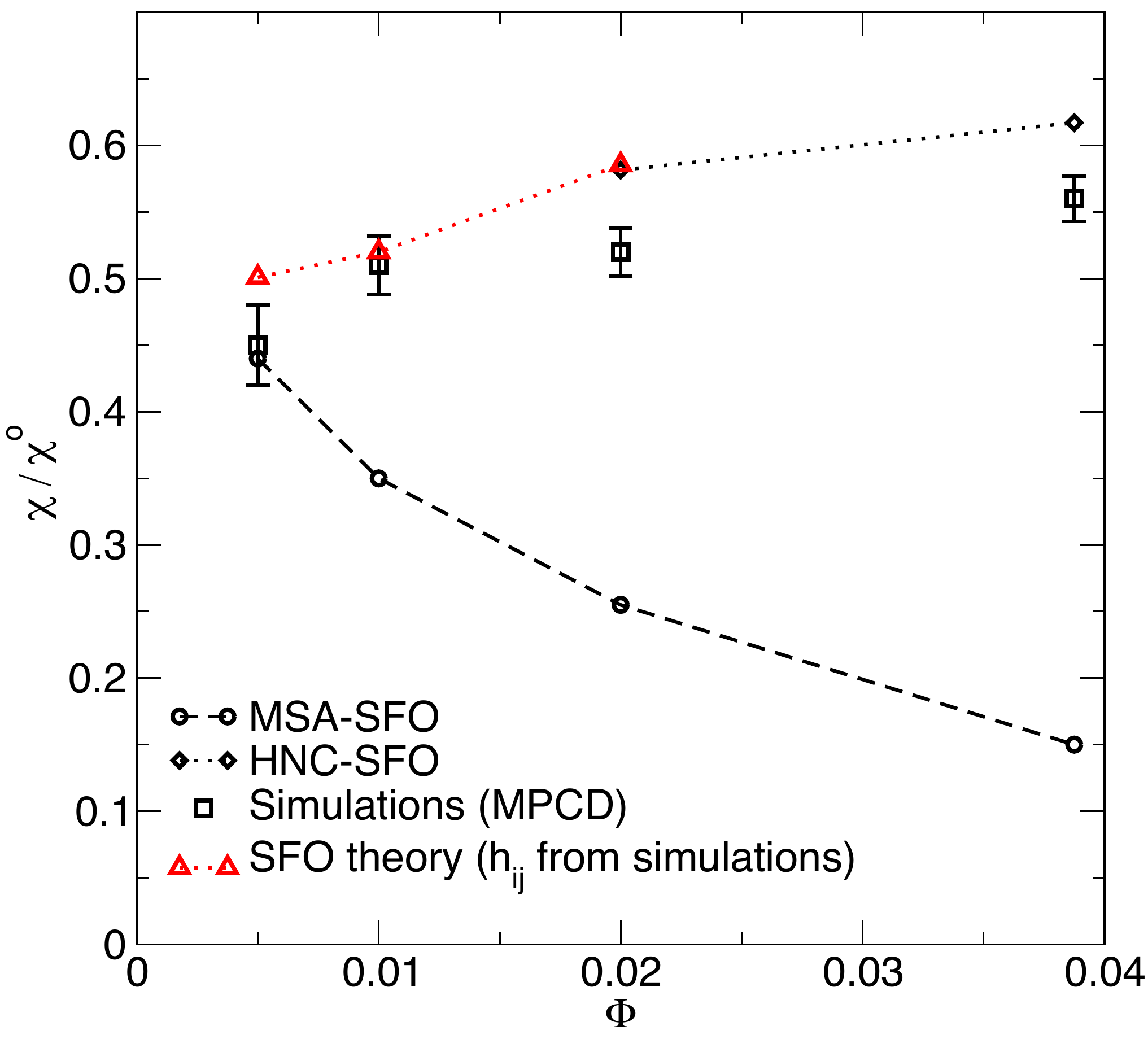}
  \caption{Electrical conductivity $\chi$ of a $1-8$ electrolyte, divided by the value at infinite dilution $\chi^ { \circ }$, as a function of the volume fraction in nanoparticles $\Phi$. The nanoparticles have a diameter of $1.05$ nm. Open squares: MPCD simulations ; Fuoss-Onsager transport theory, with total equilibrium pair correlation function $h_{ij}^{\circ}(\bf{r})$ computed with MPCD simulations (triangles), HNC (diamonds), and MSA (circles).}
  \label{figure5}
\end{figure}

\begin{figure}[h]
\centering
  \includegraphics[scale=0.38]{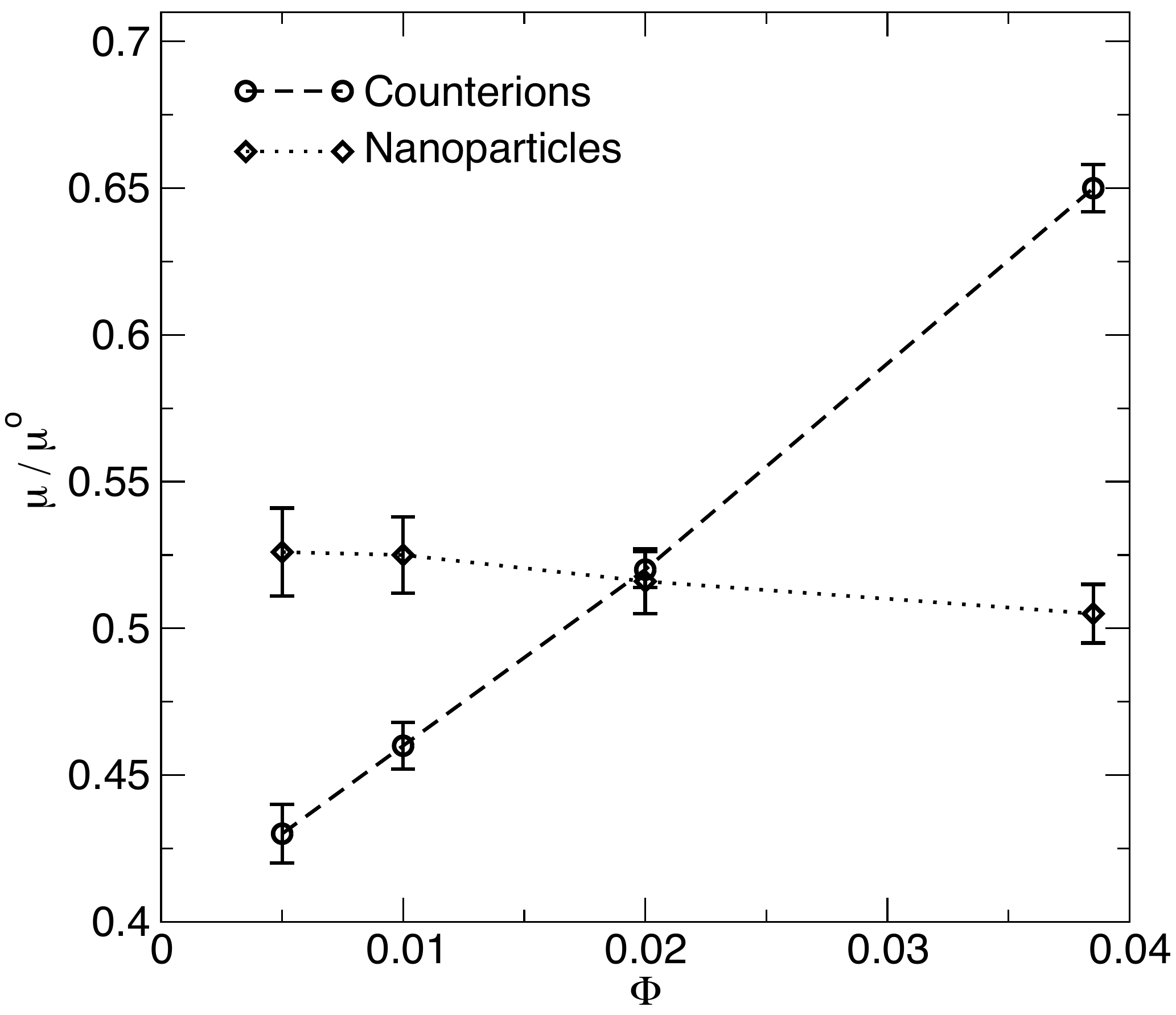}
  \caption{Electrophoretic mobilities $\mu$ of charged particles of a  $1-8$ electrolyte, divided by the value at infinite dilution $\mu ^ { \circ }$, as a function of the volume fraction in nanoparticle $\Phi$. The nanoparticles have a diameter of $1.05$ nm. The values have been computed from MPCD simulations.}
  \label{figure6}
\end{figure}

The electrical conductivity of the aqueous solution of $1-8$ electrolyte as a function of the volume fraction is shown in Fig.\ref{figure5}.
Interestingly, in the studied range of concentrations, the simulations predict that the ratio of the conductivity over the ideal conductivity $\chi/\chi^{\circ}$ increases when the electrolyte concentration increases. In the dilute limit, this is always the opposite, $\chi/\chi^{\circ}$ decreases when the electrolyte concentration increases, because in this range the corrective electrostatic relaxation and hydrodynamic forces become more intense when the concentration increases. It is noteworthy that the SFO transport theory predicts the same behavior, when one uses as input the accurate static correlation functions given either by HNC or by the simulations. On the contrary, the use of MSA closure equation predicts a monotonic decrease of $\chi/\chi^{\circ}$. This is an example for which a bad description of the equilibrium distribution functions is a clear cause of an apparent failure of the transport theory. Nevertheless, all approximations made by the transport theory, such as the use of the approximate Oseen tensor, do not prevent the prediction of the right qualitative behavior when combined with accurate equilibrium functions. 
How can we understand the increase of $\chi/\chi^{\circ}$ with the concentration predicted by both simulation and theory ? To get further insight into this, we computed the electrophoretic mobility of the nanoparticles and of their counterions. As shown in Fig. \ref{figure6}, when the electrolyte concentration increases, the mobility of the nanoparticle decreases, but the mobility of the counterions increases. We suggest that the increase of the system concentration, by decreasing the mean distance between the nanoparticles and decreasing the potential of mean force between counterions and nanoparticles, favors hoping motions of condensed counterions from one nanoparticle to another. This phenomenon would increase the mobility of counterions and the overall conductivity, while it would not affect significantly the mobility of nanoparticles.

\subsection{Failure of electrolyte theories for more asymmetric systems}

\begin{figure}[h]
\centering
  \includegraphics[scale=0.38]{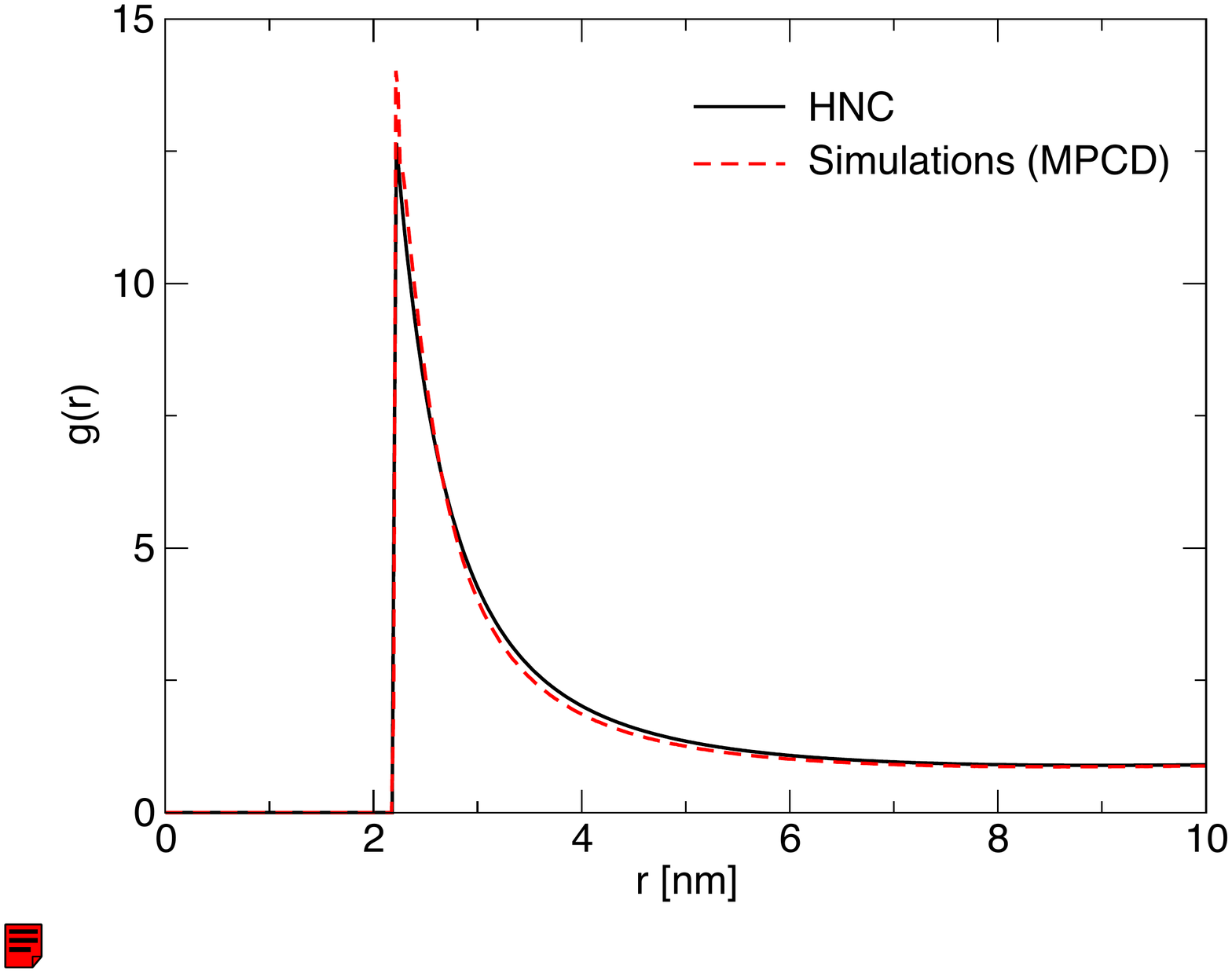}
  \caption{Pair radial distribution function g(r) between the nanoparticles and the counterions, in the case where the electrolyte is highly asymmetric ($Z_{NP}=16$ for a diameter $\sigma_{NP}=4.1$ nm), for a volume fraction in nanoparticles $\Phi=0.01$. The black curve corresponds to the results of integral equations (HNC closure), while the red dashed line corresponds to the results of numerical simulations (MPCD).}
  \label{figure7}
\end{figure}

As we have seen so far, ions of diameter close to one nanometer can be considered as {\em big ions}, even if they are highly charged. Indeed, electrolyte transport theories provide reliable estimates of the dynamic quantities, even when these nanoparticles are highly crowded and highly charged. Is it also the case for larger nanoparticles ? 
We consider the case of suspensions containing nanoparticles of diameter $\sigma_{NP}=4.1$~nm, and charge equal to $-16$e. This particle size roughly corresponds to charged micelles, such as micelles of  tetradecyltrimetyl bromide in water\cite{DurandVidalJPC06}. 
The radial distribution functions obtained with HNC and from the MPCD simulations are shown in Fig.\ref{figure7}, for the volume fraction in nanoparticle  $\phi=0.01$. We find again an excellent agreement between the theory and the simulations. 

\begin{figure}[h]
\centering
  \includegraphics[scale=0.38]{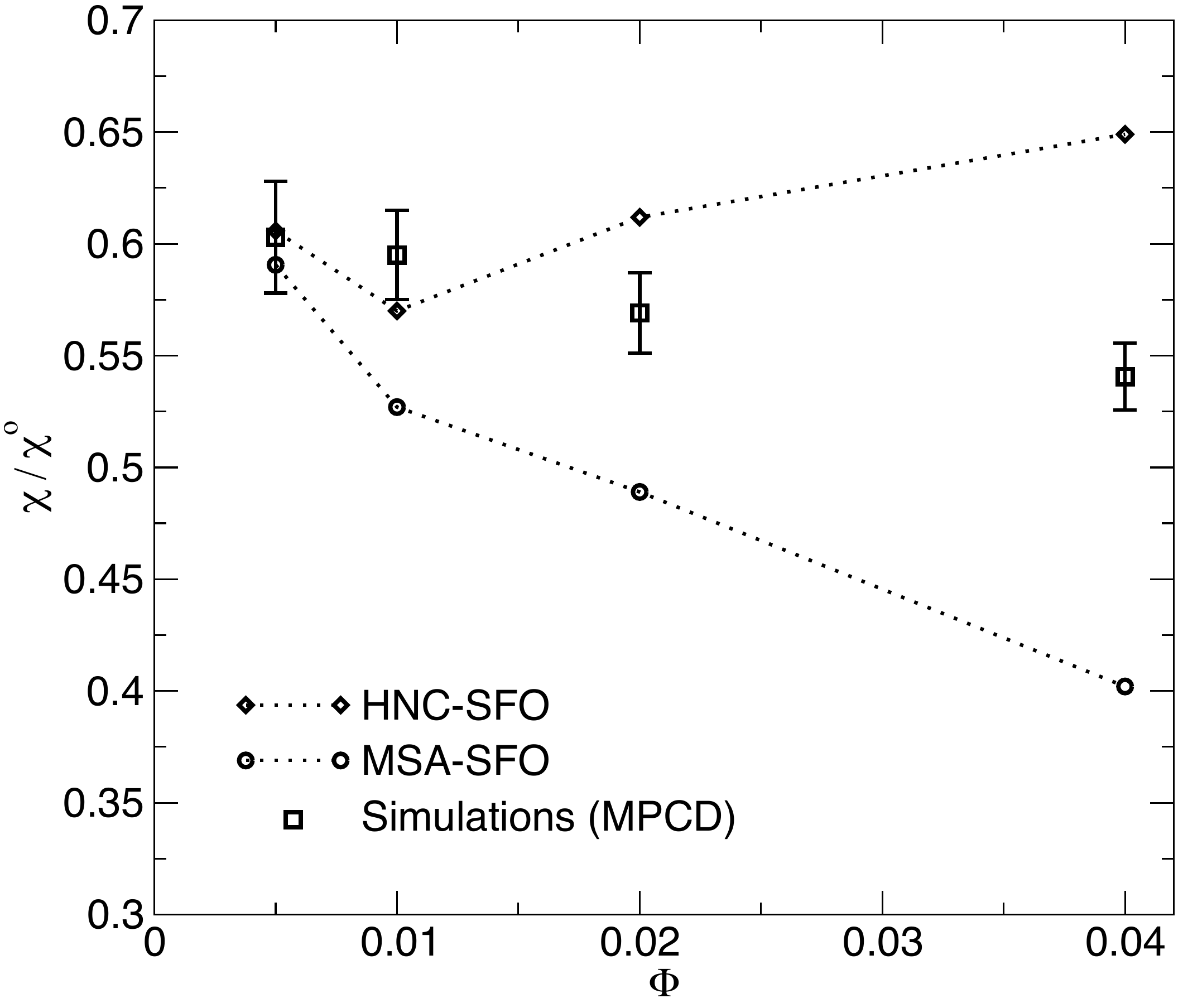}
  \caption{Electrical conductivity $\chi$ of a $1-16$ electrolyte, divided by the value at infinite dilution $\chi ^ { \circ }$, as a function of the volume fraction in nanoparticle $\Phi$. Open squares: MPCD simulations ; Open circles and diamonds: Self-consistent Fuoss-Onsager transport theory, with the HNC closure (diamonds), and with the MSA (circles), with the same parameters as MPCD.}
  \label{figure8}
\end{figure}

The electrical conductivity of the solution divided by the ideal value obtained from numerical simulation and from the SFO theories is shown in Fig. \ref{figure8}. For the most dilute system, $\phi=0.005$, all results are close to each other. Nevertheless, for all the more concentrated systems, we observe strong discrepancies between (i) both implementations of the theory, SFO-MSA and SFO-HNC, and (ii) the theories and the simulation results.   There are two possible explanations for this observation: Either the electrolyte theories are not valid for this kind of system, where the size asymmetry between the nanoparticle and its counterion is large, or the results of the simulations are not precise enough. The later explanation does not seem likely: In the concentration range investigated here, the number of solutes in the simulation box is high so that the simulations results do not suffer from low statistics. Moreover, we have shown in a previous article that the simulation methodology chosen here, with mixed coupling schemes, is quantitatively equivalent to more sophisticated coupling schemes\cite{Zhao2016}. Also, for the latter methodologies, there is an important series of work showing their ability to get the right transport properties \cite{PaddingPRE06, Imperio2011, Hecht05}. We therefore think that our investigation reveals a clear limitation of the electrolyte transport theories as they are usually implemented.  Actually, the electrophoretic mobilities predicted by the HNC-SFO calculation for the $1-16$ electrolyte are too large compared to those predicted by simulations. Moreover, these mobilities increase in the high volume fraction regime contrarily to the simulation results.  The failure of the   transport theory in this case might be due to the calculation of the relaxation force  that only relies on the calculation of the electrostatic force (see eq. \ref{eq7}). An osmotic contribution to the force coming from the large amount of counterions in the vicinity of the nanoparticle could perhaps be added to the computation of the relaxation force.   Indeed, interestingly, the results from the previous section show that charge asymmetry seems to be less of a problem than size asymmetry. 
Eventually, as far as electrolyte theory is concerned, micelles are not big ions. The equilibrium functions predicted by the HNC closure are correct, but the description of the dynamics is not.  Unfortunately, such nanoparticles can neither be considered as small colloidal particles as far as conductivity is concerned, since a straightforward calculation shows that their contribution to the conductivity is of the same order of magnitude as that of counterions, which is not compatible with the time scale separation inherent to the theories of colloidal systems. More investigations would be needed to explore the origin of the breakdown of the electrolyte transport theory and to suggest improvements of these theories. 

\section{Conclusions}

Electrolyte science have been widely developed during the twentieth century, from the emergence of the field with the Debye-H\"uckel theory to the investigation of ion-specific effects of various origins. In the meantime, theories adapted to the colloidal scale have been continuously improved, so that experimental structural and dynamical quantities could be used to extract the characteristics of the colloidal particles, including the parameters that govern the interparticle potential of mean force, such as the zeta potential. Many systems are in the limit of asymmetric electrolyte and small nanocolloidal system. For instance, polyoxometalate anions are nanoparticles of about one nanometer of diameter, and they are used as standard systems to calibrate experimental devices used to characterize colloidal suspensions. It is however more natural to consider such systems as electrolytes, as the charge of the nanoparticle is only $-4e$. In particular, the contribution of the nanoparticles to the electrical conductivity is not negligible, conversely to the case of colloidal suspensions. 

So far, the extension of electrolyte theories to describe the transport properties of suspensions of charged nanoparticles has never been properly questioned. The advances in simulation methodologies at this scale offer a way to tackle these issues. In the present study, MPCD simulations enabled to challenge the ability of Debye-Fuoss-Onsager theories to predict the electrical conductivity of suspensions of charged nanoparticles, in the salt free case. When this transport theory is combined to integral equations at the HNC level to describe the structure of the suspension, we found that the theory is able to reproduce the simulation results for nanoparticles of diameter $1$~nm, even if their charge is strong and that part of the counterions are strongly attracted by the nanoparticle. Strikingly, the theory is able to capture the nonmonotonic variation of the ratio of the conductivity over the ideal conductivity $\chi / \chi^{\circ}$. A simpler treatment of the structure, when the closure relationship of the integral equations is described by the MSA approximation, fails to describe conductivity, in particular for packing fraction higher than $0.5 \%$. 

The simulations also show that the current versions of electrolyte theories fail to describe suspensions containing larger nanoparticles ({\em e. g} of diameter $4$~nm). This indicates that in this regime, where the electrolyte is more similar to a colloidal suspension, the approximations within electrolyte transport equations are not adapted. More theoretical work is needed.

\bibliographystyle{rsc}
\bibliography{rsc}


\end{document}